\documentclass[journal]{IEEEtran}
\usepackage{hyperref}

\ifCLASSINFOpdf
\usepackage{graphicx}
\else
   \usepackage[dvips]{graphicx}
\fi
\usepackage[cmex10]{amsmath}
\usepackage{multirow}
\usepackage{color}

\begin{document}

%
\title{Distributed or Monolithic?\\A Computational Architecture Decision Framework}
%
%
%

\author{Mohsen~Mosleh,
        Kia~Dalili,
        and~Babak~Heydari
\thanks{
Mohsen Mosleh and Babak Heydari are with the School of Systems and Enterprises, Stevens Institute of Technology, Hoboken,
NJ, 07093 USA (e-mail: mmosleh@stevens.edu, bheydari@stevens.edu).}
\thanks{Kia Dalili was with the School of Systems and Enterprises, Stevens Institute of Technology, Hoboken,
NJ, 07093 USA. He is now with Facebook Inc., New York, NY 10017.}
\thanks{This work was supported by Defense Advanced Research Projects Agency/NASA contract NNA11AB35C. } }

%
%

\markboth{To APPEAR IN IEEE SYSTEMS JOURNAL, DOI: \href{dx.doi.org/10.1109/JSYST.2016.2594290} {10.1109/JSYST.2016.2594290}}%
{Shell \MakeLowercase{\textit{et al.}}: Bare Demo of IEEEtran.cls for Journals}
%



\maketitle

\begin{abstract}
Distributed architectures have become ubiquitous in many complex technical and socio-technical systems because of their role in improving uncertainty management, accommodating multiple stakeholders, and increasing scalability and evolvability. This departure from monolithic architectures provides a system with more flexibility and robustness in response to uncertainties that it may confront during its lifetime. Distributed architecture does not provide benefits only, as it can increase cost and complexity of the system and result in potential instabilities. The mechanisms behind this trade-off, however, are analogous to those of the widely-studied transition from integrated to modular architectures. In this paper, we use a conceptual decision framework that unifies modularity and distributed architecture on a five-stage systems architecture spectrum. We add an extensive computational layer to the framework and explain how this can enhance decision making about the level of modularity of the architecture. We then apply it to a simplified demonstration of the Defense Advanced Research Projects Agency (DARPA) fractionated satellite program. Through simulation, we calculate the net value that is gained (or lost) by migrating from a monolithic architecture to a distributed architecture and show how this value changes as a function of uncertainties in the environment and various system parameters. Additionally, we use Value at Risk as a measure for the risk of losing the value of distributed architecture, given its inherent uncertainty.

\end{abstract}


%
\IEEEpeerreviewmaketitle

\section{Introduction}
%
%
%
%

\IEEEPARstart{F}{or} many Engineering Systems, dealing with a growing level of uncertainty results in an increase in systems complexity and a host of new challenges in design and architecting such systems.
These systems need to respond to a set of changes in the market, technology, regulatory landscape, and budget availability. Changes in these factors are unknown to the systems architect not only during the design phase, but also during earlier phases, such as concept development and requirements analysis, of the system's life cycle \cite{heydari2012optimal}. The ability to deal with a high level of uncertainty translates into higher \textit{architecture flexibility} in engineering systems which enables the system to respond to variations more rapidly, with less cost, or less impact on the system effectiveness.

Distributed architecture is a common approach to increase system flexibility and responsiveness. In a distributed architecture, subsystems are often physically separated and exchange resources through standard interfaces. Advances in networking technology, together with increasing system flexibility requirements, has made distributed architecture a ubiquitous theme in many complex technical systems. Examples can be seen in many engineering systems: Distributed Generation, which is an approach to employ numerous small-scale decentralized technologies to produce electricity close to the end users of power, as opposed to the use of few large-scale monolithic and centralized power plants \cite{DG}; Wireless Sensor Networks, in which spatially distributed autonomous sensors collect data and cooperatively pass information to a main location \cite{yick2008wireless}; and Fractionated Satellites, in which a group of small-scale, distributed, free-flying satellites are designed to accomplish the same goal as the single large-scale monolithic satellite \cite{brown2009value}.

The trend towards distributed architectures is not limited to technical systems, and can be observed in many social and socio-technical systems, such as Open Source Software Development \cite{kogut2001open}, in which widely dispersed developers contribute collaboratively to source code, and Human-based Computation (a.k.a. Distributed Thinking), in which systems of computers and large numbers
of humans work together in order to solve problems
that could not be solved by either computers or humans
alone \cite{quinn2009taxonomy}. Despite the differences between the applications of these systems, the underlying forces that drive systems from monolithic, in which all subsystems are located in a single physical unit, to distributed architectures, consisting of multiple remote physical units, have some fundamental factors in common. For all these systems, distributed architecture enhances uncertainty management through increased systems flexibility and resilience, as well as enabling scalability and evolvability \cite{tanenbaum2002distributed}.

In spite of the growing trend toward distributed architecture, studies concerning the systems-level driving forces and cost/benefit analysis of moving from monolithic to distributed architecture have remained scarce\footnote{One attempt is \cite{crawley2015systems}. However, the authors did not quantify costs/benefits of transition from monolithic to distributed architecture based on systems-level driving forces.}. These studies are essential to decision models which determine the net value of migrating to distributed schemes. As we will argue in this paper and have shown in our previous work \cite{heydari2014}, the fundamental systemic driving forces and trade-offs of moving from monolithic to distributed architecture are essentially similar to those for moving from integrated to modular architectures. In both of these two dichotomies, increased uncertainty, often in the environment, is one of the key contributors for pushing a system toward more \textit{decentralized} scheme of architecture, in which subsystems are loosely coupled. For example, consider a processing unit. Depending on the relative rate of change and uncertainty in the use, technology upgrade or budget, the CPU can be an integrated part of the system (e.g., Smart phone), becomes modular at discretion of the user (e.g., PC), transitions to client-server architecture to accommodate smoother response to technology upgrade, security threads, or computational demand, or migrate to a fully flexible system with dynamic resource-sharing (e.g., Cloud computing).

In this paper, we formulate these problems under the umbrella of the general concept of \textit{modularity}. Modularity has often been recognized as a general set of principles---as opposed to a mere design technique---that enhance managing complex products and organizational systems. Modularity in the broadest sense of word, is defined as a mechanism to break up a complex system into discrete pieces that can then interact with one another through standardized interfaces \cite{Langlois2002}. This broad definition of modularity requires us to think of modularity as a continuous spectrum that includes a wide range of architectures which covers integrated, modular yet monolithic, and distributed schemes. This framework can be used as a basis for computational methods for deciding about systems architecture and flexibility calculations related to modularity. We will use this broad definition of modularity together with the notion of a modularity spectrum to create a framework that can be used as a basis for computational architecture decision methods and flexibility evaluation for systems.

Engineers have long held the intuition that more decentralized schemes---i.e. higher levels of modularity---increase a system's flexibility \cite{suh1990principles, ulrich1995product}. They have used modularity for complexity management in many domains, such as software \cite{sullivan2001structure}, hardware architecture \cite{baldwin2000design}, the automotive industry \cite{pandremenos2009modularity}, production networks \cite{sturgeon2002modular}, outsourcing  \cite{mikkola2003modularity}, and mass customization \cite{liu2012optimal}.
Furthermore, modularity has widely been studied and applied in organizational design and systems architecture \cite{schilling2001use}; it has been argued that a loosely coupled firm, in which each unit can function autonomously and concurrently, can benefit from increased \textit{strategic flexibility} to respond to environmental changes, due to reduced difficulty of adaptive coordination \cite{sanchez2002modularity}. 
Proper use of modularity is also argued to bring economies of scale, increase feasibility of product/component change, increase product variety, and enhance product diagnosis, maintenance, repair, and disposal \cite{huang1998modularity}. Finally, modularity is shown to help with increasing systems flexibility and evolvability by reducing the cost of change and upgrade in the system on the one hand, and facilitating product innovation on the other hand \cite{baldwin2000design}.


Despite these advantages of modularity, there are studies that show that many systems follow an opposite path toward more integration, which suggest thinking about its downsides and the underlying trade-offs \cite{schilling2000toward,sharman2004characterizing,Langlois2002,holtta2007degree}. When discussing such trade-offs, it is crucial to remember that modularity is not a binary property but a continuum, representing the degree of coupling between components of a system, and describes the extent to which a system's components can be separated and recombined \cite{schilling2000toward}. The appropriate level of modularity is, among other factors such as technical design requirements, determined by the flexibility required for the system to deal with the changes and uncertainties that the system confronts during its lifetime. The value that modularity, as a systems mechanism for managing uncertainty, adds to the system has a diminishing return. Although a low degree of modularity hampers response to environmental changes, over-modularity increases the overall cost of the system, gives rise to a host of potential problems at the interfaces. 
%
Hence, finding the appropriate level of modularity, corresponding to underlying forces in the system's environment, is a crucial step in decision-making under uncertainty from the perspective of system architecture.


In this paper, we consider distributed architecture as a part of a modular architecture spectrum and analyze the trade-offs associated with migrating from a monolithic to a distributed architecture for a real, yet simplified, case of a satellite system that was a part of a demonstration for a DARPA/NASA program on fractionated satellites. We use a conceptual framework, developed in our previous work \cite{heydari2014}, to enhance architecture decisions according to the level of composition or modularity for systems under uncertainty. We add an extensive computational layer to the framework, apply it in the context of space systems, and analyze the value of design alternatives for various uncertainty parameters and subsystems configurations. The proposed framework helps to identify design alternatives based on the level of responsiveness to environment uncertainty achievable by various levels of modularity. This provides decision makers with systemic intuition when selecting and evaluating alternatives for a given set of environment uncertainty parameters in the otherwise intractable space of design alternative possibilities. In order to determine the value of moving toward more decentralized schemes, we quantify the net gain in the value of the system that incorporates increased flexibility, and the associated cost of adopting higher levels of modularity in the system. To compute this value, we add a stochastic simulation layer on top of the proposed model that determines the conditions under which transition toward a distributed architecture is sensible.
Results of this framework are calculated in the form of probability distributions of the net value of an architectural change, to accommodate different decision makers with heterogeneous expected costs or tolerance for risk. 



The organization of the rest of the paper is as follows. In Section~\ref{Methodology}, we describe the underlying conceptual systems architecture framework \cite{heydari2014} that is used in this paper, and explain how it unifies modularity and distributed architecture on a five-stage spectrum and helps in selecting/evaluating design alternatives. In Section~\ref{CaseStudy}, we introduce the challenge of decision making for distributed architectures in space systems. In Section~\ref{FractionationValue}, we add a computational layer to the conceptual framework and build a mathematical model for a simplified case of satellite systems. Finally, in Section~\ref{Simulation}, we illustrate and compare configuration alternatives for different values of uncertainties in the environment and various system parameters, and apply different measures, such as Value at Risk (VaR), for comparison. 
\begin{figure*}[!t]
\centering
\includegraphics[width=5in]{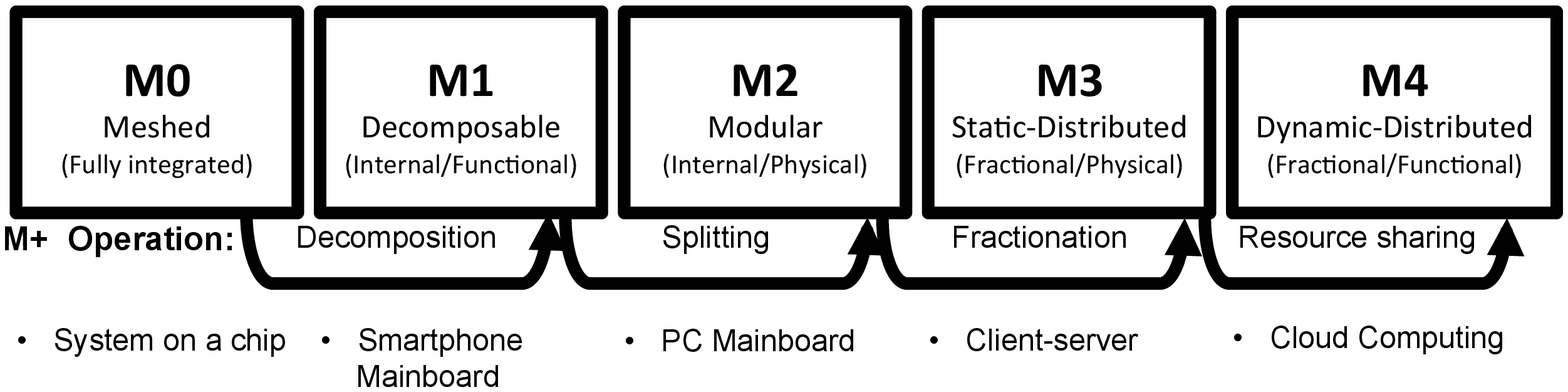}
\caption{Five-stage modularity and distributed architecture spectrum, and corresponding four M+ operations. Adaptability of the system increases as we move right. Each operation assesses the aggregate value of moving one step for a given system, subsystem or component \cite{heydari2014}.}
\label{modularity_spectrum}
\end{figure*}


\section{Modularity and distributed architecture Framework}
\label{Methodology}

In conventional systems engineering approaches, design alternatives are evaluated based on meeting requirements that are derived from a set of probable scenarios, practical and technical limitations, and cost considerations. Requirement-driven approaches, by their very nature, are insufficient for assessing a system's responsiveness to uncertainty. Unless non-functional requirements are explicitly quantified \cite{ross2015towards}, one cannot use requirement-driven approaches to compare the goodness of a more expensive design that exhibits smaller cost growth when it faces an undesired uncertainty, vis-a-vis a less expensive, yet inflexible design \cite{brown2009value}.


In this paper we use a value-centric design approach, which has been suggested as a way to overcome the shortcomings of traditional approaches and to evaluate the flexibility of design alternatives \cite{collopy2011value,collopy2002value,collopy2001economic}. In this approach, the focus is shifted from a requirement-centric to a value-centric perspective in which the designer compares the system net present value for different design alternatives over the lifetime. The system value may encompass costs and revenue streams as well as the (real) option value of flexibility resulting from modularity, scalablity, and evolvability \cite{brown2007system}.

Methods for deciding about systems architecture under uncertainty fall into two broad categories: Those that provide a set of qualitative systemic intuition to select plausible design alternatives (e.g., \cite{de2007classification,fricke2005design}), and those that take a fixed set of alternatives as input and determine their \textit{optimality} based on some exact methods (e.g., \cite{silver2007time,baldwin2000design}). In principle, one can combine these two methods to first find a set of design alternatives based on systemic intuitions and then select the most appropriate one (see \cite{selva2010integrated} for an example in space systems). However, this two-step process is not sufficient for many problems where multiple iterations between these two steps are needed. Under these circumstances, a unified framework that inherently includes both of these steps can significantly enhance the iterative process of nominating and selecting design alternatives.


As a unified framework, we employ a value-based decision making framework developed in our previous work \cite{heydari2014}. This framework is based on a systems architecture spectrum that covers a wide range of modularity/distributed architecture in complex systems and classifies the degree of modularity into five stages, as shown in Fig.~\ref{modularity_spectrum}. While in this paper we only evaluate transitions between stages in the framework, within each stage, a continuous measure of modularity can be defined depending on the specific problem and resolution needed, e.g., measures based on DSM for Modular ($M_2$) systems  \cite{holtta2007degree} and network modularity index \cite{newman2004finding} for Dynamic-Distributed ($M_4$) systems. The framework enhances systems architecture decisions under uncertainty by limiting the search space to possible alternatives with various degrees of responsiveness to uncertainty. Selecting design alternatives from different defined stages in the architecture spectrum, together with quantifying the value of the transition from one stage to another, effectively reduces the complexity of the design decision problem and provides intuitions to systems architects.

The systems architecture framework \cite{heydari2014} considers five stages of modularity, indicated by $M_0$ to $M_4$. Stage $M_0$ describes fully integrated systems, in which components are connected to each other in a way that neither physical nor functional sub-parts can be identified, e.g., System on a Chip, in which several electronic systems are integrated into a single chip. $M_0$ is considered  the minimum level of modularity and is the baseline in the modularity framework. $M_1$ represents systems with identifiable sub-parts, each responsible for a specific share of the overall system's functionality. Components at this stage, although modular in function, cannot easily be customized, replaced, or upgraded during later stages of the system's lifecycle. Smartphone and tablet mainboards can be considered to be at this modularity stage.

At stage $M_2$, similar to $M_1$, functionalities can be broken down and attributed to components. However, the related components are connected to the rest of the system via standardized interfaces. These standard interfaces allow the components to be replaced or upgraded without disrupting the rest of the system. As opposed to Smartphone mainboards, personal computers’ mainboards are at stage $M_2$, which allows users to customize or upgrade components such as memory and CPUs. While $M_0$, $M_1$ and $M_2$ encompass all cases of modularization for monolithic systems---systems comprised of a single physical unit---$M_3$ and $M_4$ cover systems with more than one unit, in which communications between units is a possibility. 

At $M_3$, certain functionalities of the otherwise monolithic system are transferred to different, and often remote, physical units. Here, we refer to these units as systems \textit{fractions}. A function can be centralized in one or more fractions, creating a client-server system in which a certain task can be delegated by a majority of fractions that lack a certain functionality, to a fraction with a powerful version of that module. A communication channel is needed for pre-processed and post-processed resources between the client units and the server fraction. Client-server computation systems are an example of systems at stage $M_3$. Unlike $M_3$, in which the division of labor among subsystems is static, in $M_4$, the task can be dynamically distributed among various fractions that have different capabilities in terms of the required functionality. Cloud computing systems are an example of systems at $M_4$. The architecture of systems at $M_4$ can be represented by complex networks models where nodes represent subsystems and edges represent interactions between the subsystems (subsystems' heterogeneous parameters can be modeled by nodes' attributes and heterogeneity of subsystems' interactions can be modeled by weighted links and multi-layered networks) \cite{heydari2015efficient,6637051}.

Moving to distributed architectures (i.e., $M_3$ and $M_4$) creates a more adaptable system, but this extra adaptability and flexibility come at a cost and, under certain conditions, may result in instability. The reason behind the increased chance of instability is that distributed architectures require complex task coordination schemes to allocate resources under uncertain demand and there are often many paths for resource exchange as well as multiple feedback loops between the components in a system with high levels of modularity (i.e., Static-Distributed). Moreover, the chance of instability increases further in systems with even higher levels of modularity (i.e., Dynamic-Distributed), in which components have a level of autonomy and their goals are not necessarily aligned with that of the whole system. 

Four \textit{M+ operations} are defined in the framework that represent transitions from one stage of modularity to the next in terms of required changes in the system architecture and the increased degree of modularity. In order to identify the optimal modularity, we have to quantify and compare the value of the system prior to the operation, to the value of the system afterward. Such evaluation requires knowledge of the system and its environment. The value of the system at each modularity level can be calculated via any of the standard system evaluation methods (e.g., scenario analysis, discounted cash flow analysis) and should consider technical, economic and life cycle parameters. 

It is worth noting that decisions in the proposed framework are based on the aggregate economic value that modular architecture can add to a system by increasing the responsiveness to environment uncertainty. Hence, the framework can be used to evaluate design alternatives that are technically viable, given factors such as physical constraints or performance requirements. The proposed model can complement the context dependent deterministic approaches for deciding about modular architecture.

\section{Case Study: Monolithic and distributed architectures in space systems}
\label{CaseStudy}

Space systems are often required to deal with a large set of uncertainties throughout their lifecycles, which in turn makes design decisions challenging and, in many cases, intractable, i.e., a large number of possible design alternatives should be evaluated against myriad number of uncertainty scenarios. There are various sources of uncertainty for space systems including technology evolution, demand fluctuation, launch failure, funding availability, and changes in stakeholders' requirements.
Accommodating most of these uncertainties in conventional monolithic designs would result in increased cost and complexity. For example, in response to components failure, the conventional approach either suggests extreme measures of reliability or the use of redundant parts, both of which result in higher mass, cost, and in most cases higher power. Addressing other types of uncertainties, such as changes in technology or stakeholder requirements, is either not possible, or requires unconventional and often costly methods \cite{saleh2003flexibility,brown2002value}.


As a result of these problems, systemic flexibility is needed to deal with the increased levels of uncertainty. New methods have been suggested, based on flexible and adaptable design, that enable spacecraft systems to respond to uncertainty more rapidly and at a reasonable cost. One approach is to deploy a constellation progressively, commencing with a small and affordable capacity, which can be increased, as needed, in stages by launching additional satellites and re-configuring the existing constellation in orbit \cite{de2003enhancing}. Another approach is to provide on-orbit servicing, which makes various options, such as service for life extension or upgrade, available after the spacecraft has been deployed \cite{saleh2003flexibility}. However, physical access to a space system is very expensive. Instead, software upgrades and changes of function can be performed through information access to the space system, providing some level of flexibility to address uncertainty \cite{nilchiani2005measuring}. Another approach to increase space systems' responsiveness is fractionated satellites, a design concept in which modules are placed into separate fractions that communicate wirelessly to deliver the capability of the original monolithic system \cite{brown2007system}.

Due to the inherent flexibility that comes with distributed and networked architectures, fractionated spacecraft are considered a viable solution for accommodating uncertainty in space systems. This is a departure from large, expensive, and monolithic satellite systems to a \textit{network} of small-scale and less expensive free-flying satellites that communicate wirelessly. In this architecture, a new fraction can be launched to become part of the network of satellites without disruption of the rest of the system. This option enables incremental development and deployment, and increases system responsiveness \cite{mathieu2005assessing,brown2002value}. 

To decide about the level of flexibility of space systems architecture---e.g., through fractionation---one has to consider the level of uncertainty and changes in the environment together with the cost associated with responding to them. Fractionated architecture---as a flexible architecture---does not come without cost and is not always the best choice for space systems. If applied inappropriately, fractionation may add no value, increase the cost and complexity, and cause instability due to multiple paths and feedback loops between fractions. Hence, one has to calculate and compare the system value over its lifetime for different design alternatives and balance it against the potential costs of each alternative, taking into account the effects of design variables and environment uncertainty parameters, to find the optimal architecture.

\begin{figure}
\centering
\includegraphics[width=8cm,height=6.3cm,keepaspectratio]{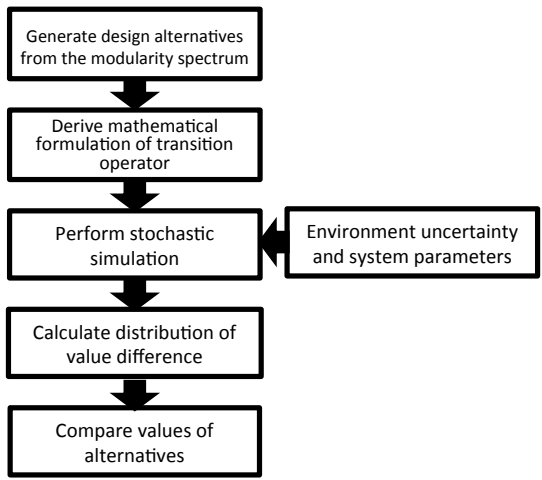}
\caption{Steps for generating and evaluating design alternatives}
\label{fig:simulationsteps}
\end{figure}

Several studies have investigated the value proposition of fractionated architecture in space systems \cite{o2011assessing,mathieu2005assessing}. However, decisions about the level of fractionation are challenging and depend on a wide range of system components' characteristics and the environment uncertainty parameters. On the one hand, some studies have investigated the systems-level trade-offs in the transition from monolithic to distributed architecture for space systems  \cite{crawley2015systems}. These studies have not operationalized the systemic trade-offs in quantifying the value that is gained (or lost) in the transition from one level of flexibility to another at the component-level. On the other hand, a wealth of studies have developed quantitative frameworks to perform tradespace exploration that often generate (enumerate) a large number of design alternatives and then evaluate them against a range of scenarios \cite{mccormick2009analyzing,maciuca2009modular,lafleur2009exploring,cornford2012evaluating,cornford2013evaluating}. Without considering the systemic trade-offs (i.e., those of moving from monolithic to distributed architecture) in the selection and evaluation of design alternatives, these quantitative frameworks do not scale well for a large number of possible design alternatives and uncertainty scenarios.  A unified framework can bridge two approaches to more effectively select and evaluate design alternatives.

In this paper, by developing computational aspects of the modularity and distributed architecture spectrum explained in Section~\ref{Methodology}, we combine the systemic intuition (i.e., how modularity can improve system responsiveness to uncertainty and what are the associated advantages/disadvantages) in selecting design alternatives with their quantitative evaluation. This approach can enhance decision-making about systems architecture by providing insights about the transition from one level of system flexibility to another and aids in selecting and evaluation of different alternatives. The proposed approach can complement existing methodologies for assessment of the value of flexibility in systems architecture design under uncertainty (e.g., \cite{collopy2011value,de2011flexibility,hazelrigg1998framework}) by integrating the role of modularity in improving the system flexibility  into the model. For example, the method suggested in \cite{de2011flexibility} is based on a four-phase procedure (estimating the distribution of future possibilities, identifying candidate flexibilities, evaluating and choosing flexible designs, and implementing flexibility), yet it does not consider the level of modularity (composition) of design alternatives neither in identifying the candidates nor in evaluating them.

We apply this approach to a case of space systems in which monolithic satellite systems and fractionated satellites can be, respectively, considered to be at levels of modularity $M_2$ and $M_3$ in the architecture spectrum inspired by a general notion of modularity. Hence, we can apply the operation $M_2 \rightarrow M_3$ for calculating the value of distributed architecture and graph it against different parameters. To accommodate stakeholders' risk tolerance, we use Value at Risk as a measure for the risk of losing the value of distributed architecture. In this case study, we particularly compare monolithic ($M_2$) and fractionated ($M_3$) architecture. However, the proposed framework can also be used to evaluate the transition from fractionated to dynamic-distributed architecture (e.g., Federated satellites \cite{golkar2015federated} and Earth observation sensor web \cite{di2010earth}) in space systems using the operation $M_3 \rightarrow M_4$.

The steps we took in calculating the value of distributed architecture are given in Fig.~\ref{fig:simulationsteps}. First, we generate design alternatives based on the proposed modularity spectrum, i.e., $M_2$ and $M_3$. Next, we derive the mathematical formulation of the transition operator. In our case we calculate the probability distribution of replacements for subsystems as well as the associated costs for each design alternative. Next, we run a stochastic simulation based on the environment uncertainty and the system parameters to find the probability distribution of the value of the transition. Finally, we compare and evaluate design alternatives based on economic measures such as Value at Risk. 

 \begin{figure*}[!t]
\centering
\includegraphics[width=6in]{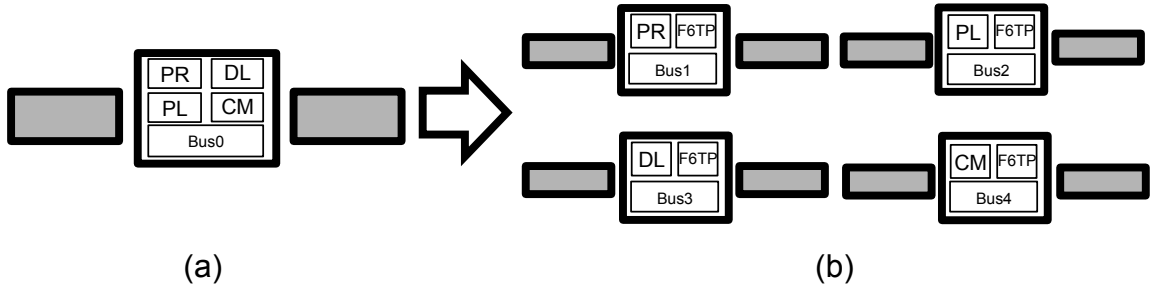}
\caption{(a) Monolithic satellite system (b) Fractionated satellites system, PR: Processor, DL: Downlink, CM: Communications Module, F6: F6TP, PL: Payload \cite{mosleh2014optimal}.}
\label{satellite}
\end{figure*}

\section{Computing Value of Distributed Architecture}

\label{FractionationValue}
In this section, we calculate the value of distributed architecture and illustrate it for a particular case in satellite systems, based on a simplified variation of fractionated architecture developed as part of System F6 \cite{DarpaF6,mosleh2014optimal}. We assume a satellite system that processes the data collected by a sensor payload and transmits them to earth via a high-speed downlink. Moreover, a data connection link from Earth can be established for maintenance purposes. In a conventional design, all of the subsystems are integrated and have to be launched together. However, in the fractionated design, subsystems can be separated in flying fractions and launched independently. The conventional monolithic system is at level $M_2$ and the fractionated system is at level $M_3$ in the five-stage architecture spectrum introduced in Fig.~\ref{modularity_spectrum}.

Subsystems communicate internally through the spacecraft bus, which also supplies power to the subsystems. The type (and cost) of a spacecraft bus is determined by the total mass of the subsystems it supports. Fractions communicate through an extra Tech Package (F6TP, in the case of System F6) that enables wireless communications among fractions. The subsystems involved in our analysis are as follows: (1) Payload: a sensor, (2) Processor: a high performance computing unit, (3) Downlink: a high-speed downlink for transmitting data to earth, (4) Communications module: a broadband access to a ground network through Inmarsat I-4 GEO constellation, (5) Bus: spacecraft bus that accommodates subsystems on board, and (6) F6TP: F6 Tech Package that enables the communications between the fractions while flying in formation.

Fig.~\ref{satellite} shows the conceptual arrangement of subsystems for the monolithic system and a possible allocation of subsystems for a fractionated configuration. In Fig.~\ref{satellite}a, all of the subsystems are integrated and communicate internally through the bus without requiring an F6TP. In Fig.~\ref{satellite}b, however, each of the four main subsystems is located in a separate fraction, along with a bus and an F6TP. 

\subsection{Value of distributed architecture}
\begin{table}
\caption{Component cost, mass, and reliability parameters. $\alpha$ is the scale parameter and $\beta$ is the shape parameter of the Weibull distribution that represents subsystems reliability (Mon.: Monolithic, Frac.: Fractionated, PR: Processor, DL: Downlink, CM: Communications Module, F6: F6TP, PL: Payload)}
\label{parameters_graph_1}
\centering
\begin{tabular}{lllll}
\hline
Component& $\alpha$& $\beta$ & Cost(K\$)& Mass(kg)\\
\hline

Payload &	15&	1.7&	27,000&	50\\
Communications&870&1.7&35,000&70\\
Downlink&190&1.7&40,000&10\\
Processor&90&1.7&30,000&20\\
F6TP&	600&	1.7&	2,000&	5\\
Bus (Mon.) &108&	1.7&	34,000&	260\\
PL Bus (Frac.) &108&1.7&28,000&180\\
CM Bus (Frac.)&108&1.7&29,000&200\\
DL Bus (Frac.)&108&1.7&25,000&150\\
PR Bus (Frac.) &108&1.7&26,000&160\\
\hline
\end{tabular}
\end{table}

In the analysis of this case, we compare the value of systems with different architectures, composed of similar main subsystems, that fulfill the same overarching goal. Moreover, we assume the systems have to function at an acceptable level of performance until the end of the project lifetime. Hence, one way to quantify the value difference between two systems is to compare the cost of designing, building, launching and operating them over a fixed project lifetime. Particularly, in our case, we calculate the value of distributed architecture as the difference in cost of running a fractionated system versus the monolithic system. Note that the value of distributed architecture could potentially be much higher than what is obtained through the method presented here, as distributed architectures also enable scalability and evolvability in the long run. As a result, this method is a conservative approach, especially in the long run and the net value calculated here can be considered as a lower bound for a given set of system parameters.

Since we assumed that the system has to function to the end of the project lifetime, we can calculate the cost of running a system as the total cost of building and launching its fractions in the beginning of the project and the cost of replacing them,  given uncertainty over the lifetime. We do not consider component costs that are identical for both architectures, e.g., subsystem design cost. We also assume that the cost of building and the mass of a fraction are equal to the sum of its subsystem costs of building and masses, respectively. Additionally, we assume a linear launch cost, proportional to the total mass of the fraction.

It is worth noting that the case study is intended to show how the proposed framework can be applied to a real-world system and present the trade-offs of distributed versus monolithic architecture for different possible distributed architecture designs. To this end, we used typical values for space systems and made few simplistic assumptions.  For example, we did not consider the integration cost, however, this can be integrated into the model as an additional cost for each subsystem, which results in a shift in the value curve of design alternatives. Additionally, we assumed linear launch cost and did not consider common costs in comparing the value of two systems. Although these could substantially change the numerical results of our illustrative case, the model can be easily extended to take more realistic assumptions to study a particular real-world system.

\subsection{Modeling uncertainty}
\label{Modeling uncertainty}

We take into account technological obsolescence and subsystem failure as two uncertainties that may affect the system over the lifetime. Other uncertainties can be integrated into our model in a similar way. We assume that the uncertainties have known probability distributions that can be approximated from historical data. We classify launch failure and in-orbit collisions as bus failure so the failure of each subsystem will be only attributed to its own reliability parameters and we assume that subsystem failure times are independent. In accordance with \cite{brown2007system}, we use the Weibull probability distribution for subsystem failure with probability characteristics presented in Table~\ref{parameters_graph_1}. For technological obsolescence, we use a Log-normal distribution and assume subsystem obsolescence times are independent. We assume that a subsystem has to be replaced when it fails or becomes obsolete in the sense that a new technology is introduced and the subsystem has to be upgraded through replacement in order to receive the benefits of the new technology. We do not consider factors such as market competition, demand, and  performance level in the decision for replacement of obsolete subsystems.

We calculate the probability density function (PDF) of replacement time for subsystems as follows. Suppose obsolescence time and failure time of subsystem $i$ are given by the random variables $O_i$ and $F_i$, respectively, and the random variable $T_i$ denotes time of replacement. The PDF, $g_i(t)$, and CDF (cumulative distribution function), $G_i(t)$, of replacement time for subsystem $i$ can be calculated as follows:
\begin{equation} \label{eq:FractionCDF}
\begin{split}
G_i(t)&=p(T_i \le t)=1-p(T_i>t) \\
&=1-p(O_i>t,F_i>t)\\
&=1-p(O_i>t)p(F_i>t)\\
&=1-(1-\Phi_i(t))(1-\Psi_i(t)).\\
\end{split}
\end{equation}
Differentiating both sides of Eq.~\ref{eq:FractionCDF} yields the following:
\begin{equation} \label{eq:FractionPDF}
g_i (t)=\varphi_i (t)(1-\Psi_i (t))+\psi_i (t)(1-\Phi_i (t))
\end{equation}
\noindent
where $\Phi$ and $\varphi $ are the CDF and PDF for obsolescence time, respectively, and $\Psi$ and $\psi$ are the CDF and PDF for failure time, respectively.
\subsection{Value at Risk}
\label{Value at Risk}
In this study, we incorporate the perspective of a single stakeholder and the risk-taking thresholds. Given the uncertainty of the input parameters, the deterministic presentation of each alternative value (e.g., expected value) is not sufficient for decision makers. A decision maker might prefer a higher cost solution with a lower risk to a lower cost solution with a higher risk. Hence, the result of the model will be presented in the form of the probability distribution of the net value of changes of a distributed architecture compared to its equivalent monolithic architecture.

As a measure for risk in evaluating alternatives, we use Value at Risk (VaR), a commonly-used measure for the risk of loss of an uncertain value in financial risk management, as well as in many non-financial applications \cite{mcneil2015quantitative}. For a given uncertain value, time horizon, and probability $p$, the 100$p$\% VaR is defined as a threshold loss measure, such that the probability that the loss over the given time horizon exceeds this figure is $p$ \cite{jorion2007value}.

For each comparison of alternatives, we obtain the distribution of the system's value over the given lifetime. Next, we find a threshold below which the area under the distribution curve represents 100$p$\% of the whole area under the curve. Here, $p$ is the probability that the fractionation value falls below the threshold.

\section{Results and Discussions}
\label{Simulation}

In this section, we calculate the probability distributions of the cost of operating a system with a distributed architecture and that of an equivalent monolithic system over a given lifetime. This calculation is based on the cost of building and launching subsystems and the probability distributions of their replacement times.

In a distributed architecture, a fraction has to be deployed and launched once one of its subsystems requires replacement. Similarly, for a monolithic architecture, we can consider a system with only one fraction such that the whole system has to be replaced if one of its subsystems becomes obsolete or fails. Hence, in order to find the value of a distributed architecture, we can compare the cost imposed by each subsystem replacement in a distributed architecture with that of an equivalent monolithic architecture over the project lifetime. Note that we assume the revenue which is generated by different distributed and monolithic design alternatives is the same and we only calculate the cost of operating the system. Calculating the revenue generated by each design alternative requires modeling the market uncertainty, the feedback of the system to its market, and the system scalability.

We formulate the cost of running the system as follows. For each fraction $j$, suppose a sequence of random variables $R_{1j}, R_{2j}, \dots ,R_{nj}$ represents the time between two consecutive replacements. A new instance of a fraction $j$ has to be deployed at times $R_{0j}=0, \sum_{i=0}^1R_{ij}, \dots , \sum_{i=0}^nR_{ij}$ for the system to function without interruption until the end of its lifetime, where $n$ is the largest integer such that $\sum_{i=0}^nR_{ij} < T $, where $T$ is the project lifetime. Suppose that the cost of building and launching a new instance of a fraction $j$ is $C_{Fj}$. The cost of running a system, $C$, with $m$ fractions is the total cost of replacing its fractions, discounted to present time. 

\begin{equation} \label{eq:TotalCost}
\begin{split}
C=&C_{F1}\sum_{i=0}^n e^{-r\sum_{k=0}^iR_{k1}}+C_{F2}\sum_{i=0}^n e^{-r\sum_{k=0}^iR_{k2}}+\\
&\dots  +C_{Fm}\sum_{i=0}^n e^{-r\sum_{k=0}^iR_{km}}.
\end{split}
\end{equation}

We use simulation to calculate the cost of running two systems with different architectures under similar conditions for subsystem replacements. For each incident of replacement of a subsystem, we calculate the associated cost of each architecture. Note that the probability of replacement for subsystems that are located in a fraction are dependent. Hence, once a subsystem in a fraction is replaced, we reset the replacement times for other subsystems in that fraction. Our simulation setup is as follows. First, we sample subsystem replacement times based on their probability distribution. Next, we find the subsystem with the earliest replacement time and calculate the cost associated with its replacement in each architecture. Next, we update the replacement time for the subsystems that are affected by the replacement. We continue this for each architecture until the earliest replacement time is greater than the lifetime. Finally, for each run of the simulation, we calculate the cost of running each system and discount it to the present time. Repeating this process a large number of times yields an approximation of the costs of running a distributed architecture and the equivalent monolithic architecture.

In our simulation, we use typical values for analysis of satellite systems according to \cite{brown2007system}; however, the same simulation approach can be applied for a more specific case. The input to the simulation includes subsystem costs, masses, and failure and obsolescence probability distribution parameters. Table~\ref{parameters_graph_1} shows the estimates for costs and masses of the main subsystems (e.g., Payload, Communication module, and Downlink), F6 Tech-Package and spacecraft buses that accommodate subsystems on board. The values for spacecraft buses are based on commercially available spacecraft buses, which are chosen according to  the subsystems on board and their total masses. $\alpha$ and $\beta$ in Table~\ref{parameters_graph_1} respectively represent the scale parameter and the shape parameter of the Weibull distribution\footnote{The PDF and CDF of the Weibull distribution are as follows:

\[
    f(x)= 
\begin{cases}
    \frac{\beta}{\alpha} (\frac{x}{\alpha})^{(\beta-1)}e^{-(x/\alpha)^\beta }  & x\geq 0\\
    0            & x <0
\end{cases}
\]

\[
    F(x)= 
\begin{cases}
    1- e ^{{-(x/\alpha)^\beta } }   & x\geq 0\\
    0            & x <0
\end{cases}
\]

 } for subsystem failure. We assume a mean value of 1 year\footnote{This is based on the assumption that in modular and fractionated satellite systems, computational and sensing subsystems will most likely be silicon-based whose obsolescence follows Moore’s law.} with a standard deviation of 3 years for the obsolescence probability distribution. We also assume that buses and F6TP do not become obsolete. Moreover, we consider \$30k per kg for launch cost, which is the average cost for commercially available launch vehicles \cite{corporartion2002space}. The discount rate in our simulation is 2\%, which is based on a forecast of real interest rates from which the inflation premium has been removed and based on the economic assumptions from 2014 for 30+ year programs \cite{OMB}.  We run our simulation for a project lifetime of up to 30 years, which is well beyond the standard design lifetime of 10 years \cite{brown2007system}. All simulation results are based on 10,000 trials.


\begin{figure}[!t]
\centering
\includegraphics[width=8cm,height=6.3cm]{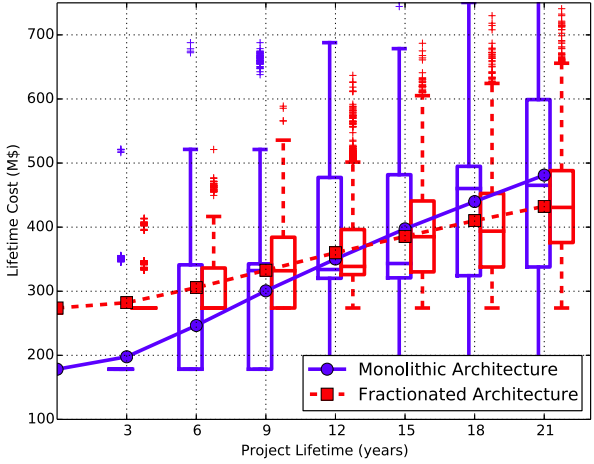}
\caption{Total cost of monolithic ($M2$) vs. fractionated ($M3$) architecture discounted to present time graphed against project lifetime for 10,000 trials. The curves represent expected values, and the boxplots depict quartiles of the cost. }
\label{fig:IntVsFrac}
\end{figure}

\subsection{Comparing fractionated and monolithic satellite systems}
\label{Comparing fractionated and monolithic Satellite Systems}

Fig.~\ref{fig:IntVsFrac} illustrates the cost of operating a fractionated satellite system, in which each main subsystem is assigned to a separate fraction, and that of a monolithic system. All costs are discounted to present time. The curves in Fig.~\ref{fig:IntVsFrac} represent the expected cost of each architecture during the project lifetime. The curves have relatively low slopes in the beginning of the system lifetime, due to the low probability of obsolescence and failure in the early years. The initial cost of running the fractionated system is greater than that of the monolithic system due to fractionation cost, i.e., the cost of building additional subsystems, such as F6TP. However, the expected lifetime cost of the monolithic system increases faster over time because the whole system must be deployed and launched once again when a subsystem fails or becomes obsolete. 

The boxplots in Fig.~\ref{fig:IntVsFrac} depict the probability distribution of cost for every 3 years. The cost variance at each point is the result of two opposing forces. On the one hand, the intrinsic property of the underlying stochastic process results in an increase of cost variance by time. On the other hand, over the project lifetime, whenever a subsystem is deployed and launched due to failure or obsolesce, the time to replace it is also reset, which reduces the cost variance. In the monolithic architecture, when a component fails, the time of replacement for the whole system is reset with a high cost. However, in the case of failure of the equivalent component in the fractionated architecture, the times to replace the other components do not change. Instead, a failure results in a lower cost. Fig.~\ref{fig:IntVsFrac} shows that the cost variance for the two systems increases by time. However, the monolithic architecture has a higher variance in every time step. This is due to the dominance of the impact of costs associated with each subsystem replacement incident.

\begin{figure}[!t]
\centering
\includegraphics[width=8cm,height=6.3cm]{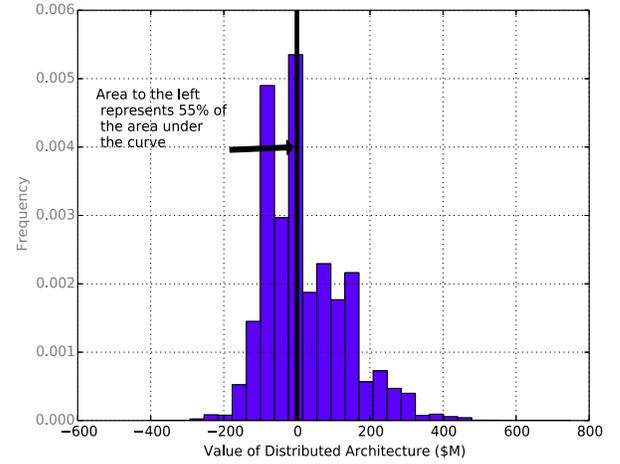}
\caption{Probability distribution for the value obtained through transition from $M_2$ (Integral) to $M_3$ (Fractional) in the modularity and distributed spectrum, i.e., project lifetime = 15 years. The vertical line at 0 represents Value at Risk 55\% is 0, meaning that area to the left represents 55\% of the total area under the curve.}
\label{fig:PayLoadDiffCost}
\end{figure}

Fig.~\ref{fig:PayLoadDiffCost} depicts the probability distribution of the value of a distributed architecture having each main subsystem in a separate fraction at $T$=15 years. As a measure of value loss under uncertainty, we use Value at Risk (VaR). As depicted in Fig.~\ref{fig:PayLoadDiffCost},  for $p=.55$ the net positive gains in transitioning from M2 to M3 is positive, meaning that there is a 0.55 probability that the value of distributed architecture falls below zero.

%
%
%

\begin{table*}[!t]
\caption{Value of Distributed Architecture for Different Payloads (EV: Expected Value, VaR: Value at Risk, $p=.25$)}
\label{Different_Payloads}
\centering
\begin{tabular}{lllllllllllllll}
\hline

\multirow{2}{*}{ } &\multicolumn{2}{c}{ }& \multicolumn{2}{c}{year 5} & \multicolumn{2}{c}{year 10} & \multicolumn{2}{c}{year 15} & \multicolumn{2}{c}{year 20}& \multicolumn{2}{c}{year 25}& \multicolumn{2}{c}{year 30}\\

&Cost(K\$)&Weight(Kg)&EV & VaR&EV & VaR&EV & VaR&EV & VaR&EV & VaR&EV & VaR\\
\hline

Payload 1&31,652&230 &-73&	-96&	-38	&-96&	-8	&-96&	20&	-85&	45&	-66&	67&	-50\\
Payload 2 &34,132&260&-66&	-96&	-21&	-96&	18&	-96&	54&	-64&	87&	-45&	118&	-26\\
Payload 3&40,332&335&-58&	-96&	4	&-96&	57&	-96&	106&	-50&	150&	-17&	190&	10.13\\

\hline
\end{tabular}
\end{table*}

\subsection{Effects of payloads' attributes on system value}
\label{Effects of payloads' attributes on system value}
In this section, we run the simulation for different payload attributes and calculate the probability distribution of the value that is obtained in transition from monolithic to distributed architecture. Table~\ref{Different_Payloads} shows the effect of payload characteristics (i.e., cost and weight) on the value of distributed architecture. For each payload, Table~\ref{Different_Payloads} shows the expected value (EV) and Value at Risk (VaR, $p$=0.25) of the value of distributed architecture every five years in the project lifetime. Payload 1, Payload 2, and Payload 3 are progressively heavier and more expensive. The results in Table~\ref{Different_Payloads} show that for a given moment in time during the project lifetime, distributed architecture creates higher value for a system with a more expensive payload and higher mass. It can be observed that a distributed architecture does not add any value to the system with Payload 1 in the first half of the lifetime (e.g., EV=-8 @year=15). However, it does add value earlier in the lifetime of the system with Payload 2 (e.g., EV=18 @year=15) and Payload 3 (e.g., EV=4 @year=10). Table~\ref{Different_Payloads} shows that, in this case, a distributed architecture is a better choice for more expensive payloads with higher masses over a shorter lifetime, all other things being equal. This is because in a monolithic system with an expensive payload, a failure in any subsystem will result in replacement of the payload, which imposes a high cost on the system. However, in the fractionated system, the payload is only replaced once a component in the payload's fraction---e.g., F6TP, bus or the payload itself---fails or becomes obsolete. Hence, when the payload is more expensive, the cost of fractionation is dominated by the savings due to elimination of unnecessary payload replacements in the fractionated architecture.


\begin{figure}[!t]
\centering
\includegraphics[width=8cm,height=6.3cm]{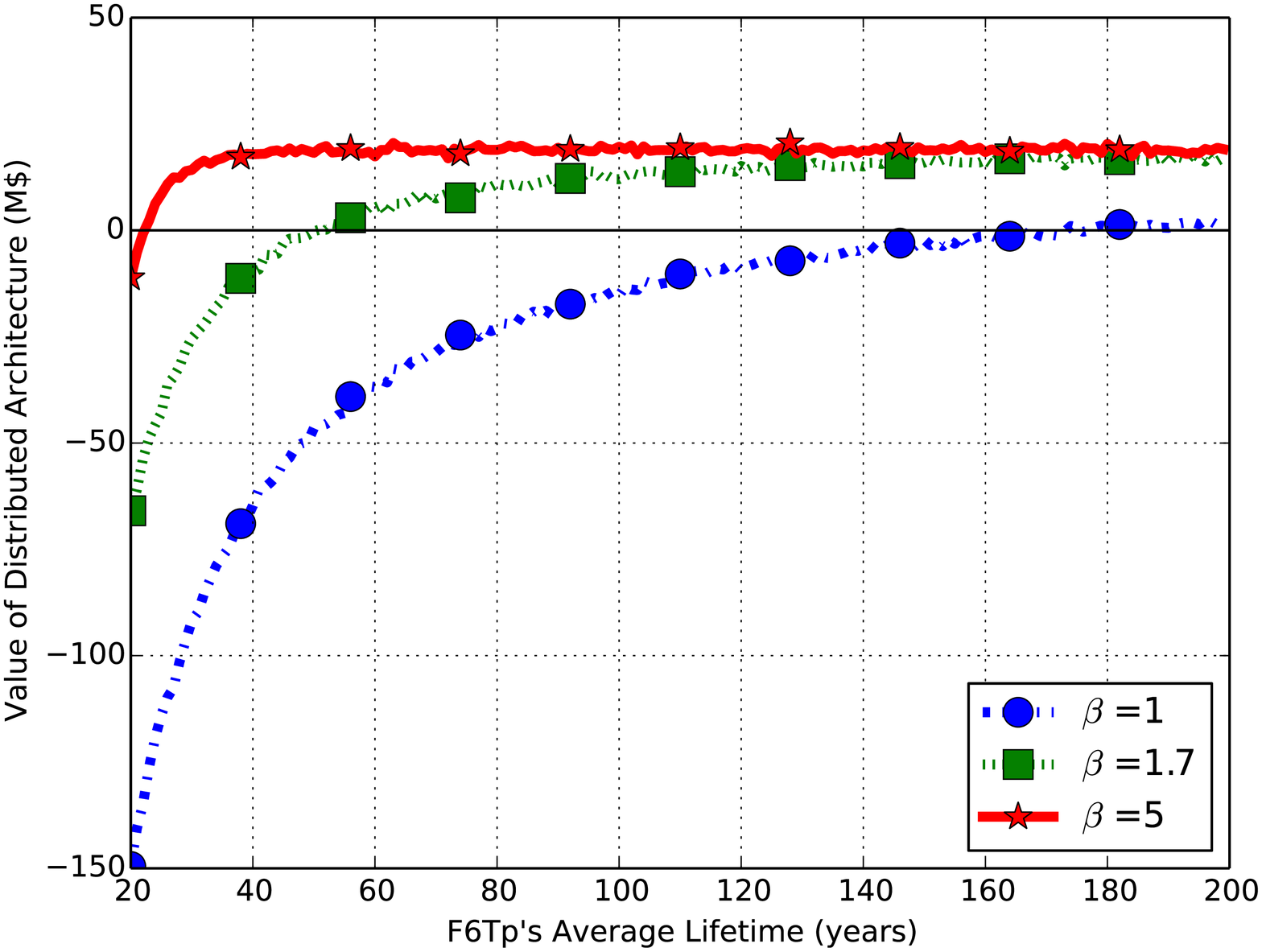}
\caption{Effects of F6TP’s reliability parameters on value of distributed architecture (i.e., project lifetime=15 years).}
\label{fig:alpha_beta}
\end{figure}


%
%
%
%
%

\subsection{Effects of F6TP's reliability parameters on system value}
\label{Effects of subsystem reliability parameters on system value}
Since F6TP is an additional subsystem required for communications between fractions in a distributed architecture, it is important to analyze how its parameters affect the value of a distributed architecture. Fig.~\ref{fig:alpha_beta} shows the effects of the F6TP reliability parameters on the value of distributed architecture for the project lifetime of 15 years. In this figure, $\beta$ is the shape parameter of the Weibull distribution that is commonly used in modeling satellite reliability. A Weibull shape parameter greater than one represents an increasing failure rate or ``wear-out". $\beta=1.7$ is a typical value in modeling reliability of satellites \cite{dezelan1999mission}. The value of distributed architecture is highly sensitive to F6TP reliability parameters for its shorter average lifetime. The curves which belong to higher values of $\beta$ represent higher increase rate of the value of distributed architecture against the F6TP's average lifetime. For a given shape parameter, when the average lifetime of the F6TP is short, there will be a huge cost to keep the system functional, due to the large number of replacements of whole fractions resulting from the F6TP failure and a monolithic architecture is always superior, as expected. Therefore, F6TP's average lifetime has to be above a certain threshold so that fractionation is sensible. Moreover, beyond a certain value of the F6TP's average lifetime (e.g., 100 years for $\beta=1.7$), the value of fractionation is not much affected by the F6TP's reliability parameters because the number of replacements of F6TP due to failure is negligible in the given lifetime (i.e., F6TP's average lifetime is much greater than the project lifetime).



\begin{table*}[!t]
\caption{Value of Distributed Architecture for Different Number of Fractions for the configurations given in Fig.~\ref{fig:con_diff_no_frac} (EV: Expected Value, VaR: Value at Risk, $p=.25$)}
\label{Different_NumberOfFractions}
\centering
\begin{tabular}{lllllllllllll}
\hline

\multirow{2}{*}{ } & \multicolumn{2}{c}{year 5} & \multicolumn{2}{c}{year 10} & \multicolumn{2}{c}{year 15}  & \multicolumn{2}{c}{year 20}& \multicolumn{2}{c}{year 25}& \multicolumn{2}{c}{year 30}\\

&EV & VaR&EV & VaR&EV & VaR&EV & VaR&EV & VaR&EV & VaR\\
\hline
Architecture (a) &-15&	-34&	11&	-34&	34&	-50&	54&	-53&	74&	-51&	92	&-46\\
Architecture (b) &-44&	-68	&-10&	-68&	18&	-68&	47&	-68&	71&	-65&	94&	-50\\
Architecture (c) &-72&	-102&	-29	&-101&	11&	-102&	45&	-73&	76&	-55&	106	&-35\\
\hline
\end{tabular}
\end{table*}

\subsection{Comparing design alternatives with different distributions of subsystems}
\label{Comparing design alternatives with different distributions of subsystems}
\begin{figure}[!t]
\centering
\includegraphics[width=9cm]{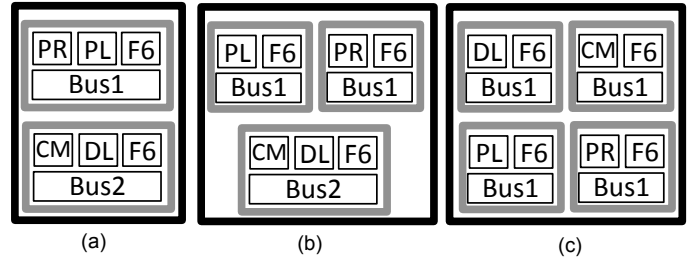}
\caption{Allocation of subsystems between fractions together with the number of fractions in Table~\ref{Different_NumberOfFractions}, (a) two fractions, (b) three fractions, (c) four fractions (PR: Processor, DL: Downlink, CM: Communication Module, F6: F6TP, PL: Payload)}
\label{fig:con_diff_no_frac}
\end{figure}

Distributed architecture (level $M_3$ in the proposed spectrum) includes a range of architectures with different numbers of fractions and allocations of subsystems. In this section, we study the effect of subsystem allocation on the value of fractionation. In our case, there are 15 ($B_4=15$ \cite{bell}) possible allocations of subsystems into fractions. We demonstrate the comparison of design alternatives for four architectures (including the monolithic architecture) with different number of fractions. However, the same analysis can be applied to all possible alternatives with the model being fairly tractable. We compare the fractionation value for the three architectures depicted in Fig.~\ref{fig:con_diff_no_frac} based on the monolithic equivalent architecture.

The architecture in Fig.~\ref{fig:con_diff_no_frac}a has two fractions. The Processor and Payload are placed in one, and the Communications module and Downlink are placed in another fraction. Fig.~\ref{fig:con_diff_no_frac}b depicts a system with three fractions. The Payload and Processor are in their own fractions, while Communications module and Downlink are located together in the same fraction. Finally, Fig.~\ref{fig:con_diff_no_frac}c represents a system in which each subsystem is placed in a separate fraction.

We calculate the probability distribution of the value of transition from monolithic to distributed architecture for each architecture independently and compare the results. The architectures in Fig.~\ref{fig:con_diff_no_frac} are progressively---from a to c---more flexible but more expensive to build. Table~\ref{Different_NumberOfFractions} shows the expected value (EV) and Value at Risk (VaR, $p$=0.25) of the distributed architecture every five years in the project lifetime for the systems conceptually depicted in Fig.~\ref{fig:con_diff_no_frac}. For each fraction in these architectures, we estimated mass and cost of the spacecraft bus based on the total mass of subsystems that are on board and according to commercially available spacecraft buses. 

Table~\ref{Different_NumberOfFractions} shows that systems with a larger number of fractions have a lower value in the earlier years of their lifetime. This is due to the dominance of high fractionation cost in the beginning of the lifetime---i.e., the higher cost of launching more fractions and the cost of additional subsystems. However, later in the system lifetime, the value of a distributed architecture in the systems with a larger number of fractions outgrows that of systems with a smaller number of fractions. This shows that the benefits of responsiveness of the system with a larger number of fractions becomes dominant over a longer lifetime. The results in Table~\ref{Different_NumberOfFractions}  demonstrate the trade-off between higher flexibility and its associated costs i.e., higher flexibility results in less cost in responding to environment uncertainty, yet making the system more flexible is costly. The results suggest that for projects with longer expected lifetime, it is worth investing on the flexibility of the system through higher number of fractions.

\section{Conclusion}
\label{Conclusion}


Building on a conceptual systems architecture framework in this paper, we developed a computational approach for decisions about the level of modularity of system architecture. We quantified the value that is gained (or lost) in the transition from a monolithic system architecture to a distributed system architecture. We applied this approach to a simplified case of a space system, 
and compared the value difference between the two architectures as a function of uncertainties in various system and environment parameters, such as cost, reliability, and technology obsolescence of different subsystems, as well as the distribution of subsystems among fractions. We used Value at Risk, a commonly-used measure for the risk of loss of an uncertain value, to accommodate the stakeholders' threshold of risk.




Distributed architectures---and in the case study of this paper, fractionated satellites---also increase scalability and resilience and foster innovation. For example, to accommodate uncertainty in the demand, additional modules can be launched throughout the lifetime of the system and become part of it. Similarly, upon failure of a module the fraction can evolve so that the system continues to function. None of these advantages were explicitly considered in assessing the value of fractionation in our model, which was focused solely on flexibility and uncertainty management. Ignoring these additional advantages, our proposed approach represents a lower bound for the value  that is obtained in the transition to distributed architecture for the given set of system parameters. We also believe that resilience mechanisms can, in theory, be modeled and integrated into the proposed framework. However, the effect of architecture on scalability needs to be modeled by considering the feedback of the system to its market, something that requires an additional layer on top of the proposed framework. Quantifying the impact of distributed architecture on creativity mechanisms is even more challenging due to a need to simultaneously integrate technological, behavioral, and economic factors into the model. Thus, it is often best to keep models of these systems features semi-qualitative and separate from flexibility models, unless the exact context and path of the system is known. This is rarely the case and often defeats the original purpose of moving toward distributed architectures in the first place.

Additionally, in this paper we used a single criteria stochastic decision model under uncertainty to find the architecture that \textit{optimally} respond to the uncertainty in the environment. Future research is needed to integrate multiple criteria and multiple stakeholders decision models into the proposed model. 



%

%

\section*{Acknowledgments}

The authors would like to thank DARPA and all the government team members for supporting this work as a part of the System F6 program. In particular, we would like to thank Paul Eremenko, Owen Brown, and Paul Collopy, who lead the F6 program and inspired us at various stages during this work. We would also like to thank Steve Cornford from JPL for his constructive feedback at various stages of this project. Our colleagues at Stevens Institute of Technology, especially Dr. Roshanak Nilchiani, were a big support of this project. We would also like to thank other members of our research group, Complex Evolving Networked Systems lab, especially Peter Ludlow, for various comments and feedback that improved this paper.

\ifCLASSOPTIONcaptionsoff
  \newpage
\fi




\bibliographystyle{IEEEtran}
\bibliography{MonolithicOrDistributed}
%



%
\newpage





\end{document}